\documentclass[11pt,a4paper]{article}
\usepackage{jheppub}
\usepackage[sort&compress]{natbib}
\usepackage{url}
\usepackage{hyperref}
\usepackage{amsfonts}% Include figure files
\usepackage{epsfig}\usepackage{dcolumn}% Align table columns on decimal point
\usepackage{bm}% bold math
\usepackage{slashed}
\graphicspath{{./figuras/}}

\def\br{\begin{eqnarray}}
\def\er{\end{eqnarray}}
\def\be{\begin{equation}}
\def\ee{\end{equation}}

\def\({\left(}
\def\){\right)}

\def\chi{\tilde{\chi}_0}

\def\lesssim{\mathrel{\hbox{\rlap{\hbox{\lower4pt\hbox{$\sim$}}}\hbox{$<$}}}}
\def\gtrsim{\mathrel{\hbox{\rlap{\hbox{\lower4pt\hbox{$\sim$}}}\hbox{$>$}}}}

\title{A Model with Chiral Quarks of Electric Charges -4/3 and 5/3}

\author[a]{Alexandre Alves,}
\affiliation[a]{Departamento de Ci\^encias Exatas e da Terra,
Universidade Federal de S\~ao Paulo, Rua Prof. Artur Riedel, 275, Diadema-SP, Brazil}
\emailAdd{aalves@unifesp.br}

\author[b]{E. Ramirez Barreto,}
\affiliation[b]{Centro de Ci\^encias Naturais e Humanas, Universidade Federal do ABC, Avenida dos Estados, 5001, Santo Andr\'e-SP, Brazil}
\emailAdd{elmerraba@gmail.com}

\author[b]{D. A. Camargo,}
\emailAdd{dacamargov@gmail.com}

\author[b]{A. G. Dias}
\emailAdd{alex.dias@ufabc.edu.br}

\abstract{
We present a new model based on the $SU(3)\otimes SU(2)\otimes U(1)$
symmetry, in which there is a new consistent set of chiral fermion
fields that renders the model free from anomalies. The new fermions do not
share the usual family structure of the Standard Model and some of
them have exotic electric charges, as the quarks $X$ and $Y$ with electric
charge $5/3$ and $-4/3$, respectively. Interestingly, the model contains a new heavy neutral lepton which may be a dark matter candidate. Two Higgs doublets are present in
our construction, so that two CP even scalars are present in the model particle spectrum. One of them is similar to the Standard Model Higgs boson, while the other one couples mainly with the new exotic fermions. We performed a discovery analysis showing that the 8 TeV LHC can find the $Y$ quark from single and  pair production with masses from $300$ GeV up to $\sim 750$ GeV. We also show that the new spectrum does not contribute significantly to the oblique EW parameters, and that dangerous flavor changing neutral currents are suppressed. Characteristic signatures from the other new fermions in the model are also commented.
}

\keywords{Beyond the Standard Model, Heavy Quark Physics}
\begin{document}
\maketitle
\flushbottom

\section{Introduction}
\label{intro}
The consistent theoretical structure of the Standard Model (SM) has been
brought to an even more confident basis with the recent discovery
of a scalar boson that closely resembles a SM Higgs boson~\cite{atlas-cms-higgs1,atlas-cms-higgs2}. As an immediate consequence, the hypothesis of existence of new chiral fermions coupling with a SM like Higgs boson has become more restricted. New colored fermions should not have large couplings with the observed
SM like Higgs once they would increase its production cross section, through gluon fusion,
by a large factor and it leads to a conflict with the observed rates
which are close to the SM predictions. A heavy fourth sequential family, for example, is
now disfavored on these grounds~\cite{djouadi-lenz-2012,kuflik-nir-volansky-2013}.

Precision electroweak data constrain the mass splitting
between the fourth family quarks, while data from
B-meson physics restrict their mixing pattern~\cite{fourthfamily1,fourthfamily2,fourthfamily3,fourthfamily4,fourthfamily5,fourthfamily6,fourthfamily7}.
Direct searches performed by the ATLAS and
CMS collaborations rule out mass-degenerate fourth generation
quarks with masses below 685 GeV~\cite{quarkfour}.
In addition, a heavy charged lepton with a mass around 100 GeV has
also been excluded at LEP2~\cite{leptonfour}.

However, there are theoretical approaches that include
new fermions within multiplets different from  the  SM sequential families.
Examples are the 3-3-1 models~\cite{331a,331b} in which some of the new particles
can be fermionic leptoquarks with electric charges $5/3$ and $-4/3$, in units of the electron charge. The LHC has a great potential to discover particles like these as shown in Ref.~\cite{abd-2012}. For more examples of quarks and leptons in different multiplets from the SM ones see Ref.~\cite{frampton1}. Within the context of the electroweak (EW) gauge symmetry group of the SM, one approach is to include new quarks fields forming vector-like doublets in such a way to evade the problem of having a strong coupling with the Higgs boson.  Previous studies show that quarks from these vector-like doublets can be discovered with collisions at the TeV energy scale~\cite{contino-servante-2008,oscar,aguilar2009,
cacciapaglia-2011,cacciapaglia-2012,exotic-leptons1,exotic-leptons2,exotic-leptons3,okada,buchkremer-et-al2013}.

Quarks with exotic electric charges have been searched in high energy colliders along the years.
For example, ATLAS and CMS, at the LHC, have placed a lower limit of $\sim 650$ GeV
for quarks with electric charge $5/3$ by considering their decays into
$W^{\pm}$ plus top quarks~\cite{D0-exquark-2007a,D0-exquark-2007b,D0-exquark-2007c}, while the D0 and CDF collaborations, at the Tevatron, have excluded a top quark with electric charge $-4/3$ at a high statistical confidence level~\cite{D0-exquark-2007a,D0-exquark-2007b,D0-exquark-2007c,exotic_charge1,exotic_charge2,cdf_charge}.

Theoretically, the introduction of new fermions can  be done straightforwardly if they are of vector-type, or in a more restricted form if they are of chiral-type, that is, with their left- and right-hand components transforming differently under the gauge group. The restriction in the last form come from the cancellation of all gauge anomalies, which specifies what set of fermions are mathematically consistent. We took this guideline and looked for a new set of chiral fermions, out of the SM sequential family, which can be shown not to contradict the existing indirect constraints and also able to be probed directly at the LHC.

Our aim in this paper is, then, to consider a new set of chiral fermions, with exotic electric charges, forming nontrivial multiplets of the SM symmetry group $SU(2)_{L}\otimes U(1)_{\mathcal{Y}}$ of the electroweak (EW) interactions.
In order to obtain quarks with exotic electric charges ($-4/3$ and $5/3$), we analyzed
the conventional electric charge operator and the corresponding hypercharge
values postulating two new doublets with the exotic quarks. One of these doublets has hypercharge $\mathcal{Y}=7/6$ and is formed by quarks $X$ and $U^\prime$, with electric charges $5/3$ and $2/3$, respectively. The other one has hypercharge $\mathcal{Y}=-5/6$ and is formed by quarks $D^\prime$ and $Y$, with electric charges $-1/3$ and $-4/3$, respectively.

For the consistency of the model, achieving anomalies cancellation, we must include two fermionic doublets containing new leptons exhibiting the same pattern, one unit of electric charge above or below the SM ones. The first of these doublets has hypercharge $\mathcal{Y}=1/2$ and is formed by the fermions $E^\prime$ and $N$ with electric charges $+1$ and zero, respectively. The second one has hypercharge $\mathcal{Y}=-3/2$ and is formed by the fermions $E$ and $F$ with electric charges $-1$ and $-2$, respectively. From these sets of new leptons, the neutral one, $N$, may be rendered stable by imposing some extra symmetry. This leads to a natural candidate for dark matter. Two Higgs doublets are also introduced for the EW symmetry breaking. All these new fermions have distinct signatures and we will discuss some of them.

Exotic quarks appear in other constructions beside the 3-3-1 models and the models with vector quarks we mentioned above, but the phenomenology of these quarks has been considered separately, that is, in models which have one or another kind of new quarks.
On the other hand, the model we are going to present here requires the appearance, by mathematical consistence, of a set of new fermions with an essential interconnection among them. Such an scheme with exotic fermions is also motivated in models where the electroweak symmetry breaking is due to strong dynamics marked by heavy quarks condensation~\cite{hill-simmons-review2003}.

We perform an analysis aiming the LHC potential to discover one of the new quarks in the model, which has electric charge $-4/3$, the $Y$ quark. We also enumerate and discuss other possible search channels, including the exciting possibility of a fermionic dark matter in the particle spectrum. The model is shown not to have conflicting impacts on the oblique EW parameters and Flavor Changing Neutral Currents (FCNC) effects. We also discuss tree level unitarity constraints on the new heavy chiral quarks.

The paper is organized as follows: in section~\ref{model} we show  the model construction and discuss FCNC effects, in sections~\ref{stu} and \ref{unit} we discuss the impact of the new fermions on the oblique parameters and on unitarity issues, respectively.
Section~\ref{signals} is devoted to the most promising searching channels at hadron colliders; in the section~\ref{pheno} we present our phenomenological analysis for the $Y$ quark at the LHC, and we draw our conclusion at the section~\ref{conclusao}.

\section{Setting of the model}
\label{model}

We now show a model construction based on the $SU(2)_{L}\otimes U(1)_{\mathcal{Y}}$ gauge group. Exotic chiral fermion fields are introduced, but we maintain the structure that left-handed fermions form $SU(2)_{L}$ doublets,  while the right-handed fermion fields are  singlets. Being the electric charge operator defined as $Q\,=I_{3}+\mathcal{Y}$, let us consider a quark doublet with hypercharge $\mathcal{Y}=7/6$, which is one unit augmented in respect to the SM ones, plus the corresponding right-handed singlets. This leads to the two exotic quarks composing the doublet having electric charges $5/3$ and $2/3$, respectively. In order to not have gauge anomalies additional multiplets of fermionic fields must be introduced. A simple solution for canceling the anomalies $ \left[SU(3)_{C}\right]^{2}\otimes U(1)_{\mathcal{Y}}$, $\left[SU(2)_{L}\right]^{2}\otimes U(1)_{\mathcal{Y}}$, and $\left[U(1)_{\mathcal{Y}}\right]^{3}$, is then to have an additional quark doublet with $\mathcal{Y}=-5/6$, plus two doublets of exotic leptons with $\mathcal{Y}=-3/2,\,1/2$. In this way, the anomaly free new matter fermionic content we set up is for the quarks
\begin{eqnarray}
\label{qex}
 &  & \mathcal{\psi}_{L}^{X}\equiv\left[\begin{array}{c}
X_{L}\\
U_{L}^{\prime}
\end{array}\right]\sim\left(\mathbf{2,\,}7/6\right),\quad X_{R}\sim\left(\mathbf{\,1,\,}5/3\right),\quad U_{R}^{\prime}\sim\left(\mathbf{1,\,}2/3\right),\cr
 &  &\cr
 &  & \psi_{L}^{Y}\equiv\left[\begin{array}{c}
D_{L}^{\prime}\\
Y_{L}
\end{array}\right]\sim\left(\mathbf{2,\,}-5/6\right),\quad D_{R}^{\prime}\sim\left(\mathbf{\,1,\,}-1/3\right),\quad Y_{R}\sim\left(\mathbf{1,\,}-4/3\right),
\end{eqnarray}
and for the leptons
\begin{eqnarray}
 &  & \Psi_{L}^{N}\equiv\left[\begin{array}{c}
E_{L}^{'+}\\
N_{L}
\end{array}\right]\sim\left(\mathbf{2,\,}1/2\right),\quad
E_{R}^{'+}\sim\left(\mathbf{1,\,}1\right),\quad N_{R}\sim\left(\mathbf{1,\,}0\right),\cr
 &  &\cr
 &  & \Psi_{L}^{F}\equiv\left[\begin{array}{c}
E_{L}^{-}\\
F_{L}^{--}
\end{array}\right]\sim\left(\mathbf{2,\,}-3/2\right),\quad E_{R}^{-}\sim\left(\mathbf{1,\,}-1\right),\quad F_{R}^{--}\sim\left(\mathbf{1,\,}-2\right),
 \label{nl2}
\end{eqnarray}
in which the numbers between parenthesis refers to transformation properties under $SU(2)_{L}$ and $U(1)_{\mathcal{Y}}$, respectively. The singlets $E_{R}^{'+}\sim\left(\mathbf{1,\,}1\right)$,  $E_{R}^{-}\sim\left(\mathbf{1,\,}-1\right)$,  and $N_{R}\sim\left(\mathbf{1,\,}0\right)$  are irrelevant for canceling the anomalies, once the first two form a vector fermion field, and the last has zero hypercharge. But they are important for mass generation and mixing among these leptons in order to provide consistent decay chain beginning with $F$, not allowing other stable charged lepton besides the electron. If $N$ is the lightest of the new leptons then with an appropriated extra symmetry such a particle could be made stable and, therefore, a dark matter candidate.

As we see, beside the $X$ quark with electric charge $5/3$, we also have a $Y$ quark  with  electric charge $-4/3$. In fact, the fermion content above extend in $\pm 1$ the range of the SM particles electric charges allowing for quarks charges $\mp\, 4/3,\,\mp \,1/3,\,\pm \,2/3,\,\pm\, 5/3$, and for leptons charges $0,\,\pm\, 1,\,\pm 2$.

We consider two Higgs doublets for breaking the EW symmetry, $H_{1,2}\sim\left(\mathbf{2,\,}1\right)$, with  vacuum expectation values $\langle H_{1,2}\rangle=\left[0\,\, v_{1,2}/\sqrt{2}\right]^{T}$. The first Higgs doublet, $H_{1}$, is assumed to have a tree level couplings to the SM quarks fields, while the second one, $H_{2}$, with the exotic quarks fields. For supporting this we take into account the existence of a  discrete symmetry $Z_4$ remaining from some physics at very high energy scale, such that the fields transforming non-trivially have charges as shown in Table \ref{zns}. The reason for this choice is that it results in a mixing of the new quark fields  $U^{\prime}$ and $D^{\prime}$   with the ordinary quark fields, which can be naturally controlled by a high energy suppression scale $\Lambda\gg v_{1,2}$. Thus, it prevents inconsistency with any established experimental fact on flavor physics. The mixing will be accomplished through nonrenormalizable operators  suppressed by the scale $\Lambda$, as we will see.

\begin{table}
\centering
\begin{tabular}{|c|c|c|c|c|c|c|}
\hline
 & $H_{2}$  & $\psi_{L}^{X}$ & $X_{R}$ & $U_{R}^{\prime}$ & $\psi_{L}^{Y}$ & $D_{R}^{\prime}$ \tabularnewline
\hline
\hline
Z$_{4}$  & $w_{1}$ & $w_{2}$ & $w_{3}$ & $w_{1}$ & $w_{1}$ & $w_{2}$ \tabularnewline
\hline
\end{tabular}
\caption{\label{zns}Charges of the fields transforming non-trivially  under $Z_{4}$, where $w_{n}\equiv e^{i\frac{\pi n}{2}}$. }
\end{table}

The tree level Yukawa Lagrangian for quarks is
\begin{eqnarray}
-\mathcal{L}_{Y} & = & y_{ij}^{u}\overline{q_{iL}}\widetilde{H}_{1}u_{jR}^{\prime}
+y_{ij}^{d}\overline{q_{iL}}H_{1}d_{jR}^{\prime}
\nonumber \\
 & + & y^{U}\overline{\psi_{L}^{X}}H_{2}U_{R}^{\prime}
 +y^{D}\overline{\psi_{L}^{Y}}\widetilde{H}_{2}D_{R}^{\prime}
 \nonumber \\
 & + & y^{X}\overline{\psi_{L}^{X}}\widetilde{H}_{2}X_{R}
 +y^{Y}\overline{\psi_{L}^{Y}}H_{2}Y_{R}
 +h.c.\label{tlyl}
\end{eqnarray}
where $q_{iL}$ are the SM quarks in left-handed
doublets, with $u_{iR}^{\prime}$ ($d_{iR}^{\prime}$) the right-handed u-type (d-type) quark singlets, respectively, as usual.
The lowest order mixing among the $U^{\prime}$ and $D^{\prime}$
fields with the $u^{\prime}$ and $d^{\prime}$ fields arises from
dimension six operators according to
\begin{equation}
-\mathcal{L}_{mix}=\frac{\left(H_{1}^{\dagger}H_{2}\right)}{\Lambda^{2}}
\left[c_{i}^{u}\,\overline{\psi_{L}^{X}}H_{2}u_{iR}^{\prime}+
c_{i}^{d}\,\overline{\psi_{L}^{Y}}\widetilde{H}_{1}d_{iR}^{\prime}\right]+h.c.
\label{Qqmix}
\end{equation}
In this Lagrangian of effective operators the coefficients $c^q_i$ are supposed
to be of order one. With the vacuum expectation values  $\langle H_{1,2}\rangle$
Eqs. (\ref{tlyl}) and (\ref{Qqmix}) give rise to a mass Lagrangian for
the quark fields in the symmetry basis
\begin{equation}
-\mathcal{L}_{mass}=\sum_{q=u,d}\mathbf{q}_{L}^{\prime}\mathcal{M}^{q}\mathbf{q}_{R}^{\prime}+
m_{X}\overline{X_{L}}X_{R}+m_{Y}\overline{Y_{L}}Y_{R}+h.c.,
\label{lmass}
\end{equation}
in which $\mathbf{q}_{L,R}^{\prime}\equiv\left(q_{1}^{\prime},
q_{2}^{\prime},q_{3}^{\prime},Q^{\prime}\right)_{L,R}^{T}$,
with $Q^{\prime}$ standing for $U^{\prime}$ or $D^{\prime}$ symmetry
eigenstates;  $\mathcal{M}^{u}$ is the mass matrix  for u-type quarks which has the following
texture,
\begin{align}
\mathcal{M}^{u} & =\left[\begin{array}{cc}
M^{u}_{ij} & 0\\
\left[x^{u}_i\right]^{T} & m_{U}
\end{array}\right]\label{mmq}
\end{align}
where
\begin{align}
M_{ij}^{u} & =y_{ij}^{u}\frac{v_{1}}{\sqrt{2}},\quad m_{U}=y^{U}\frac{v_{2}}{\sqrt{2}},\quad
x_{i}^{u}  =\frac{c_{i}^{u}}{2\sqrt{2}}\frac{v_{1}v_{2}^{2}}{\Lambda^{2}},
\label{mdef}
\end{align}
with similar form for the d-type quarks; and the mass of $X$ and $Y$ quarks given by
\begin{equation}
m_{X}=y^{X}\frac{v_{2}}{\sqrt{2}},\quad m_{Y}=y^{Y}\frac{v_{2}}{\sqrt{2}}\,.
\label{mxy}
\end{equation}

A texture as in Eq. (\ref{mmq}) was treated in \cite{cacciapaglia-2011,cacciapaglia-2012}
for a model with vector quarks $X$ and $U^\prime$. It can be seen that in our model all mixing among the exotic and ordinary quarks are naturally suppressed by $v_{1} v_{2}/\Lambda^2$ (with the presence of a scalar singlet $\phi$ this suppression turns out to be controlled by the ratio $\langle\phi\rangle/\Lambda$, see below).
In fact, being $\Lambda\gg v_{1,2}$, the matrices entries are such that $x^{u,d}_{i}\ll M^{u,d}_{ij},\, m_{U,D}$.
We show in Appendix \ref{sec:ApA} the details of the calculation for the mixing matrix elements, with their explicit suppression factor.

Although taken here to be small the mixing play the crucial role of making the exotic quarks to decay into ordinary quarks, through EW processes.  With the fields without prime referring to the mass eigenstates the charged current interactions are, at first order,
\begin{eqnarray}
& &\mathcal{L}_{Ud}=-\frac{g}{\sqrt{2}}\mathcal{V}_{4i}
\overline{U}_{L}\slashed{W}^{+}d_{iL} +h.c.
\label{ccUd}\\
& &\mathcal{L}_{uD}=-\frac{g}{\sqrt{2}}\mathcal{V}_{i4}
\overline{u_{iL}}\,\slashed{W}^{-}D_{L}+h.c.
\label{ccuD}\\
& &\mathcal{L}_{X}=-\frac{g}{\sqrt{2}}\left(\mathcal{V}_{L}^{u}\right)_{4i}
\overline{X}_{L}\slashed{W}^{+}u_{iL}
-\frac{g}{\sqrt{2}}\overline{X}_{L}\slashed{W}^{+}U_{L}+h.c.
\label{ccXu}\\
& &\mathcal{L}_{Y}=-\frac{g}{\sqrt{2}}\left(\mathcal{V}_{L}^{d}\right)_{4i}
\overline{Y}_{L}\,\slashed{W}^{-}d_{iL}
-\frac{g}{\sqrt{2}}\overline{Y}_{L}\,\slashed{W}^{-}D_{L}+h.c.
\label{ccYd}
\end{eqnarray}
The mixing matrix elements $\mathcal{V}_{4i}$, $\mathcal{V}_{i4}$, and $(\mathcal{V}_{L}^{u,d})_{4i}$, in terms of the parameters in the model are given in Appendix.

Flavor changing neutral currents processes are also predicted to occur once the doublets of new fermions have quantum numbers different from the SM ones. The mixing of  $U^\prime$ and $D^\prime$ quarks with ordinary quarks leads to non-diagonal interactions with the $Z$ boson. At first order  these interactions are
\begin{equation}
\mathcal{L}^{f.c.}=-\frac{g}{cos\theta_{W}}\left[ \left(V_{L}^{u\dagger}B_{1}^{u}\right)_{i}\overline{u_{iL}}\slashed{Z}\,U_{L}-
\left(V_{L}^{d\dagger}B_{1}^{d}\right)_{i}\overline{d_{iL}}\slashed{Z}\,D_{L}\right] +h.c.\,,\label{fcnc}
\end{equation}
in which $B_{1}^{u,d}$ has a suppression factor with the scale $\Lambda$, as we see from Eq. (\ref{b1}). The unitary matrices $V_{L}^{u,d}$ are the usual ones involved in the diagonalization of the ordinary quarks. Therefore, the flavor changing neutral currents interactions are also suppressed.

It is beyond of the scope of this work to discuss the constraints from the actual ex\-pe\-ri\-men\-tal limits on the coefficients $(V_{L}^{u,d\dagger}B_{1}^{u,d})_{i}$ in  Eq. (\ref{fcnc}). We just assume that, in this particular point, the suppression in those coefficients are high enough for avoiding tension with any established experimental fact on this issue.

Concerning the uncolored fermion fields in Eq. (\ref{nl2}) the question for their masses and mixing with ordinary  leptons could be worked out straightforwardly. But, we will not develop that here since our aim here is the exotic quark phenomenology.

The most general renormalizable scalar potential with the two doublets, and invariant under  the $Z_4$ symmetry in Table \ref{zns}  has an extra global $U(1)$ symmetry which would bring a Goldstone boson into the particle spectrum. It can be avoided in many ways, and we comment two possibilities. One option is letting the discrete symmetry to be explicitly broken in the potential by a term like $(H_1^\dagger H_1)(H_1^\dagger H_2)$. Another option is introducing a scalar field singlet $\phi\sim (\mathbf{1,\,}0)$ transforming as $\phi\rightarrow w_3\,\phi$,  so that the scalar potential would be
\begin{equation}
V(H_i,\,\phi)=  V_{\mathcal{H}}+\left(f H_{1}^{\dagger}H_{2}\phi + \lambda\,\phi^4 +h.c.\right),
\label{potesc}
\end{equation}
in which $V_{\mathcal{H}}\equiv V_{\mathcal{H}}(H_i,\,\phi)$ represents the sum of all possible Hermitian operators involving the doublets $H_i$ and the singlet $\phi$, with $f$ being a parameter with mass dimension. With a vacuum expectation value $\langle\phi\rangle=v_\phi/\sqrt2$  the Goldstone boson could be made having feeble interactions with the SM fields, and so it is harmless. In this case we still have dimension five operators for the exotic-ordinary quarks mixing,  which adds effectively to each term in Eq. (\ref{Qqmix}) by  replacing  ${(H_{1}^{\dagger}H_{2})}c^{u,d}_i/{\Lambda^{2}}\rightarrow (H_{1}^{\dagger}H_{2})c^{u,d}_i/\Lambda^{2}+\phi^*a^{u,d}_i/\Lambda$, with $a^{u,d}_i$ of the unity order. The effect on the mixing matrices is that the suppression factors are also changed according with the replacement $\frac{1}{2}\frac{v_1v_2}{\Lambda^2}c_i^{u,d}\rightarrow \frac{1}{2}\frac{v_1v_2}{\Lambda^2}c_i^{u,d}+\frac{1}{\sqrt2}\frac{v_\phi}{\Lambda}a_i^{u,d}$. If $v_\phi > v_{1,2}$, then the exotic-ordinary quarks mixing would be controlled by the ratio $\frac{v_\phi}{\Lambda}$ (see Appendix). Such  ratio can give sizable matrix elements for the mixing matrices in Eqs. (\ref{ccUd}), (\ref{ccuD}), (\ref{ccXu}), (\ref{ccYd}), and (\ref{fcnc}) even for very high $\Lambda$ if $v_\phi\lesssim \Lambda$.

\section{Oblique parameters}
\label{stu}
Before discussing the phenomenological aspects of the model we want to show that the model is consistent with the precision electroweak constraints. Indirect effects of new particles on neutral-current and observable effects involving $Z$ and $W$ bosons by means of the oblique $\left(S,T,U\right)$
parameters \cite{peskin-takeuchi} can be evaluated straightforwardly. Taking
the formulas in Ref~\cite{he-polonsky-su2001}, each pair of fermions
$\left(\psi_{1},\psi_{2}\right)$, with masses $\left(m_{1},m_{2}\right)$,
whose left-handed components form a doublet $\Psi\equiv\left(\psi_{1L}\,\,\psi_{2L}\right)^T\sim(\mathbf{2,\,}\mathcal{Y})$
of hypercharge $\mathcal{Y}$, and their right-handed components are singlets
$\psi_{1R}$, $\psi_{2R}$, gives the following contribution to the
the oblique parameters
\begin{align}
S_{\Psi} & =\frac{N_{C\psi}}{6\pi}\left[1-2\mathcal{Y}\,\mathrm{ln}\frac{x_{1}}{x_{2}}+
\frac{1+8\mathcal{Y}}{20x_{1}}+\frac{1-8\mathcal{Y}}{20x_{2}}\right],\label{ps}\\
T_{\Psi} & =\frac{N_{C}}{8\pi s_{W}^{2}c_{W}^{2}}F\left(x_{1},x_{2}\right),\label{pt}\\
U_{\Psi} & =-\frac{N_{C\psi}}{2\pi}\Bigg\{\frac{x_{1}+x_{2}}{2}
-\frac{\left(x_{1}-x_{2}\right)^{2}}{3}+\left[\frac{\left(x_{1}-x_{2}\right)^{3}}{6}
-\frac{1}{2}\frac{x_{1}^{2}+x_{2}^{2}}{x_{1}-x_{2}}\right]\mathrm{ln}\frac{x_{1}}{x_{2}}\nonumber \\
 & +\frac{x_{1}-1}{6}f\left(x_{1},x_{1}\right)+\frac{x_{2}-1}{6}f\left(x_{2},x_{2}\right)
 +\left[\frac{1}{3}-\frac{x_{1}+x_{2}}{6}-\frac{\left(x_{1}-x_{2}\right)^{2}}{6}\right]f
 \left(x_{1},x_{2}\right)\Bigg\}\label{pu}
\end{align}
in which $N_{C}=3\left(1\right)$ is the color degree of freedom of quarks
(leptons),
\[
F\left(x_{1},x_{2}\right)=\frac{x_{1}+x_{2}}{2}-\frac{x_{1}x_{2}}{x_{1}
-x_{2}}\mathrm{ln}\frac{x_{1}}{x_{2}}
\]
\[
f\left(x_{1},x_{2}\right)=\Bigg\{\begin{array}{c}
-2\sqrt{\Delta}\left[\mathrm{arctan}\frac{x_{1}-x_{2}+1}{\sqrt{\Delta}}
-\mathrm{arctan}\frac{x_{1}-x_{2}-1}{\sqrt{\Delta}}\right]\\
0\qquad\qquad\qquad\qquad\qquad\qquad\qquad\\
\sqrt{-\Delta}\,\mathrm{ln}\frac{x_{1}+x_{2}-1
+\sqrt{-\Delta}}{x_{1}+x_{2}-1-\sqrt{-\Delta}}\qquad\qquad\qquad\qquad\qquad
\end{array}\begin{array}{c}
\left(\Delta>0\right)\\
\left(\Delta=0\right)\\
\left(\Delta<0\right)
\end{array}
\]
with $x_{i}=m_{i}^{2}/M_{Z}^{2}$, and $\Delta=2\left(x_{1}+x_{2}\right)-\left(x_{1}-x_{2}\right)^{2}-1$.

For $S_{\psi}$ it is considered that $m_{1,2}^{2}\gg M_{Z}^{2}$,
which is consistent with the assumptions we take on the masses of the
new fermions. We shall restrict our analysis to the case where $\mid m_{1}-m_{2}\mid<M_{Z}$
so that we always have $\Delta>0$. It can be seen by direct application
of the above formulas that $U_{\psi}\ll S_{\psi},T_{\psi}$ for all
mass ranges we consider here, so that the $U$ parameter will be disregard.
It can be checked that the new  multiplets of fermionic fields in Eqs. (\ref{qex}) and
(\ref{nl2}) lead to corrections to
the total $S$ and $T$ parameters still inside the actual $95\%$
CL interval \cite{pdg2012} for a wide range of masses.

We observe that if $m_{E}<m_{F}$ then
the set $\psi_{L}^{F}$, $E_{R}^{-}$, $F_{R}^{--}$ in Eq. (\ref{nl2})
gives a negative contribution for $S$, offsetting positive contributions
from the other fields, keeping the parameter under control.
For example, taking $m_{E}=100$ GeV, $m_{F}=190$ GeV, $m_{E^{\prime}}=140$
GeV, $m_{N}=130$ GeV, and for the new quarks $m_{X}=350$ GeV, $m_{U}=340$
GeV, $m_{D}=325$ GeV, $m_{Y}=340$ GeV we obtain $\left(S,\, T\right)=\left(0.15,\,0.16\right)$
(in this case $U=0.02$). We have considered at this point that the neutral lepton $N$
is heavy and of Dirac type.

\begin{figure}
\centering
\includegraphics[scale=0.5]{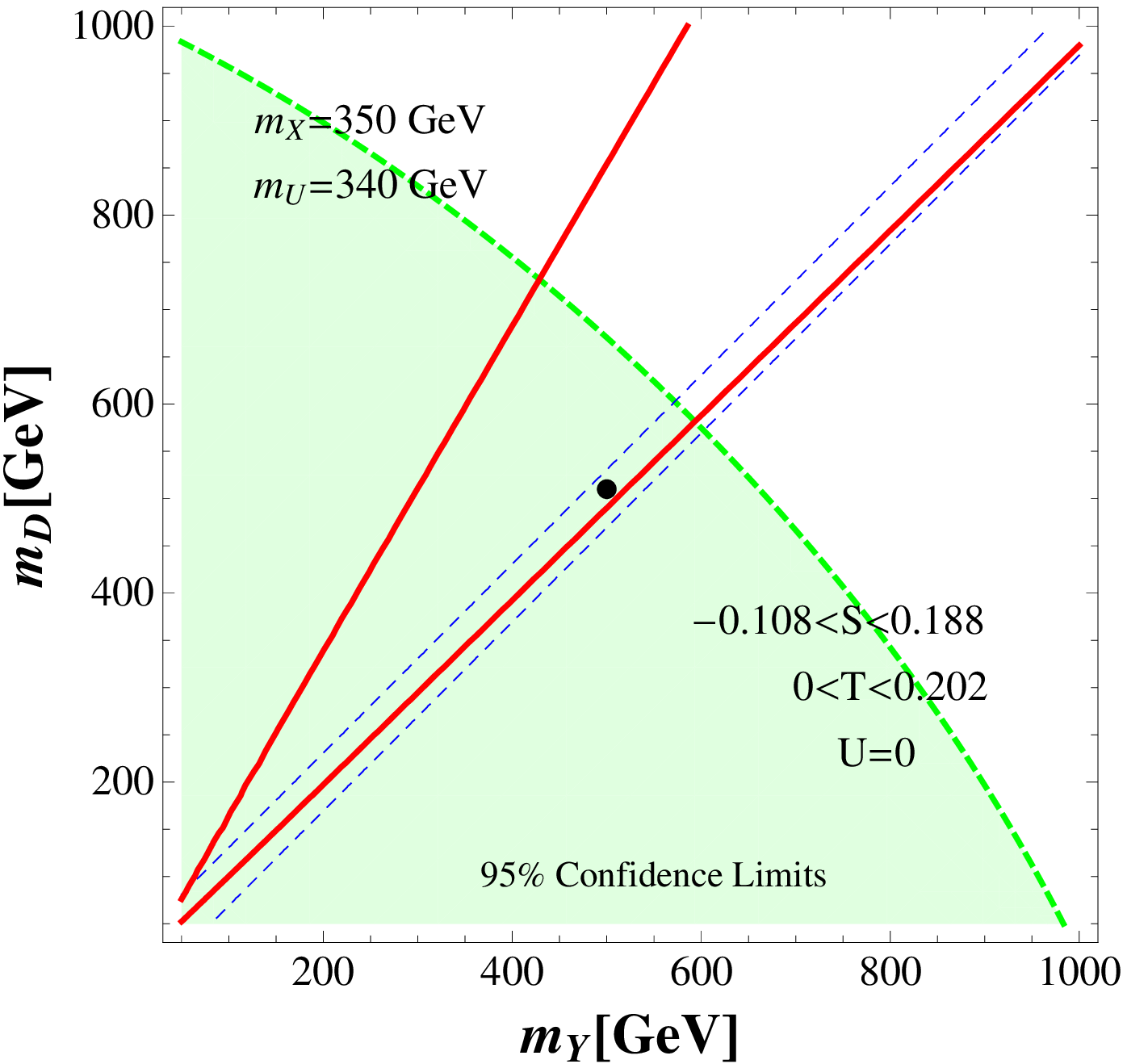}
\includegraphics[scale=0.5]{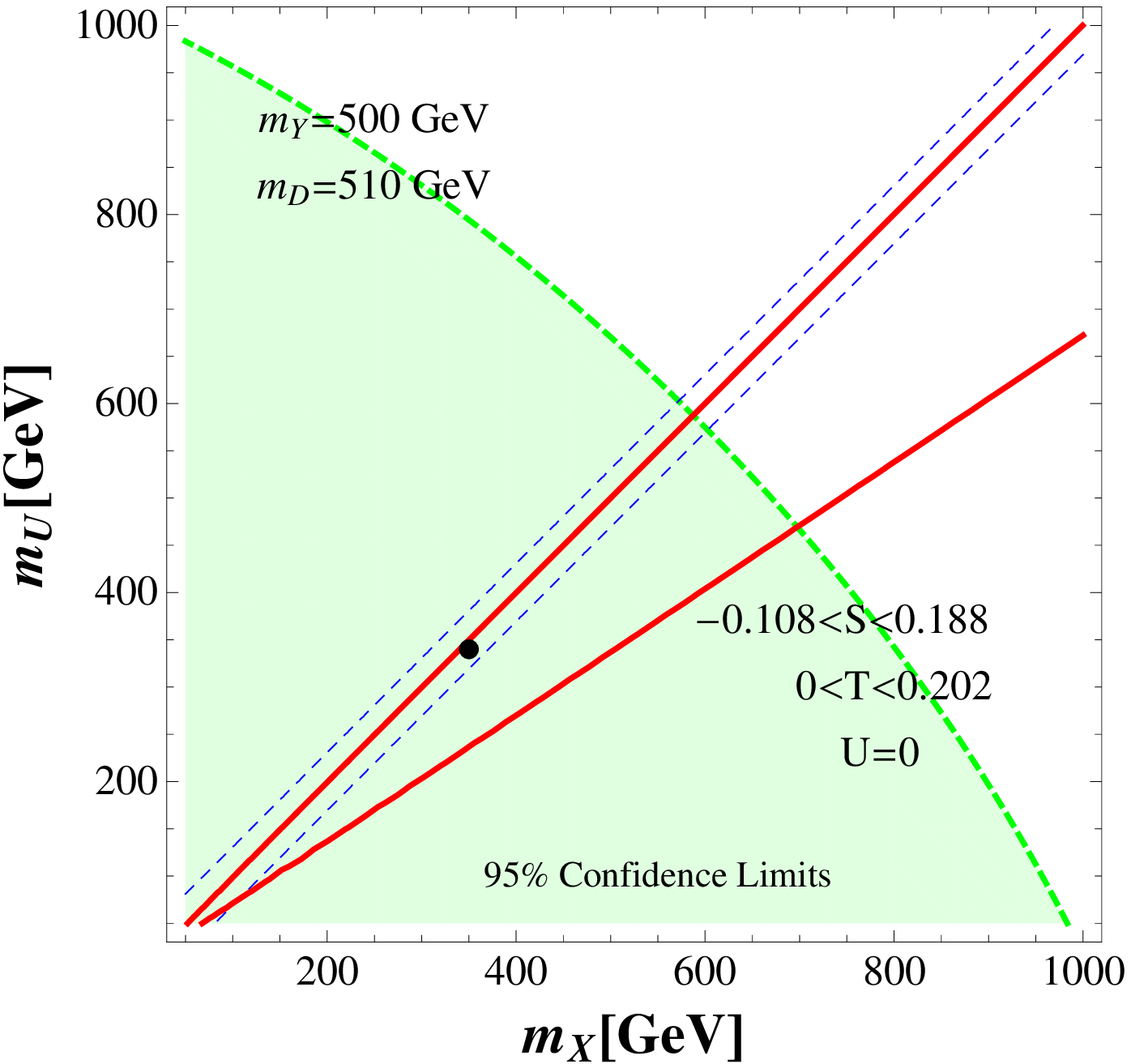}
\caption{\label{fig-stu} In the left panel we present the allowed range in the $m_{D}$ {\it versus} $m_{Y}$ plane where the $S$ and $T$ parameters lies inside the 95\% C.L. region.  In the right panel, the 95\% CL region for $S$ and $T$ but in the $m_{X}$ {\it versus} $m_{U}$ plane. The region between the solid red(dashed blue) curves represent the allowed region of the $S$($T$) parameter at this confidence level. The black dots represent a valid solution to $(S,T,U)$, within the 95\% confidence limit, for the fixed masses quoted in upper left corner of the figures. The shaded green area is the region of unitarity conservation based on the analysis of section~\ref{unit} and Eq.~\ref{m1m2}.}
\end{figure}

In the left panel of Figure \ref{fig-stu} it is shown the allowed range for $m_{D}$
and $m_{Y}$ for fixed values $m_{E}=100$ GeV, $m_{F}=190$ GeV,
$m_{E^{\prime}}=140$ GeV, $m_{N}=130$ GeV, $m_{X}=350$ GeV, $m_{U}=340$
GeV. The region between the solid red curves represent the the 95\% Confidence Level (CL) region for the $S$ parameter, and the region between the dashed blue curves the 95\% CL region allowed to the $T$ parameter.
In the right panel of Figure \ref{fig-stu} it is shown the allowed range for $m_{X}$ and
$m_{U}$ for fixed values $m_{E}=100$ GeV, $m_{F}=190$ GeV, $m_{E^{\prime}}=140$
GeV, $m_{N}=130$ GeV, $m_{D}=510$ GeV, $m_{Y}=500$ GeV, within
the 95\% CL region for $S$, and $T$ parameters and the curves have the same meaning of the previous case. The black dots represent a valid solution to $(S,T,U)$, within the established confidence limit, for the fixed masses quoted in the figure.
\section{Tree level unitarity bound on the quark masses}
\label{unit}

 The amplitudes for $Q_1\overline{Q}_1\rightarrow Q_2\overline{Q}_2$, where $Q_1$ and $Q_2$ are heavy chiral fermions, grow like $G_F m_Q^2$ in the regime of very high energies~\cite{chanowitz}. As $m_{Q_1}$ and $m_{Q_2}$ get large, the partial wave amplitudes saturate the tree level unitarity bound, the theory becomes strongly coupled and the perturbation theory becomes meaningless.

 Consider the amplitude for ${\cal M}_B(Q_1\overline{Q}_1\rightarrow Q_2\overline{Q}_2)$ at the Born approximation. The $S$-wave ($J=0$) is the leading contribution for the scattering amplitude and is given by~\cite{chanowitz,dicus,barger}
\begin{equation}
a_0=\frac{1}{32\pi}\int_{-1}^1 -i{\cal M}(Q_1\overline{Q}_1\rightarrow Q_2\overline{Q}_2) d(\cos\theta)
\label{a0}
\end{equation}

The unitarity constraint impose that $|a_0|\leq 1$ which determines an upper bound on the fermion mass. The scattering amplitude, in the case where $Q_1$ and $Q_2$ are new chiral quarks forming color-neutral channels, ${\cal M}_B(Q_1\overline{Q}_1\rightarrow Q_2\overline{Q}_2)$, receives contributions from s-channel and t-channel diagrams with a $Z$ boson, and $h_1$ and $h_2$ Higgs bosons. The Higgs boson contributions to Eq.~\ref{a0} cancel between the helicity amplitudes whereas the $Z$-boson ones lead to~\cite{chanowitz}
\begin{equation}
|a_0| = \frac{T_{3Q}^2G_F}{2\sqrt{2}\pi}\left[3(m_{Q_1}^2+m_{Q_2}^2)+\sqrt{9(m_{Q_1}^2-m_{Q_2}^2)+16m_{Q_1}^2m_{Q_2}^2}\; \right]
\label{m1m2}
\end{equation}
from which follows the most stringent bound on the mass of a fermion when $Q_2=Q_1=Q$
\begin{equation}
m_Q < \left(\frac{\pi\sqrt{2}}{5T_{3Q}^2G_F}\right)^{1/2}
\label{mq}
\end{equation}
assuming a  negligible mixing between the new quarks. We show in Figure~\ref{fig-stu} the unitarity conserving regions in the $m_X$ \emph{versus} $m_U$ and $m_Y$ \emph{versus} $m_D$ planes alongside the 95\% C.L. regions of the $STU$ parameters.

For an up(down)-type quark $T_{3Q}=\pm 1/2$ and we get an upper bound of $\sim 600$ GeV for the $X$, $Y$, $U$ and $D$ quarks. As we shall discuss, this is a mass range which is fully accessible to the 8 TeV LHC run. However, this bound is naive in the sense it does not take into account possible new contributions of a more fundamental theory that could push the unitarity violation to heavier masses. Moreover, a more intricate Higgs sector could give extra contributions to the scattering amplitudes and cancel, at least partially, the contributions of the partial $S$-wave amplitudes that grow with the new fermion mass.

Our proposed model also serve as a starting point upon which new constructions can be built. One possible direction is to extend the unitarity region for the new heavy chiral fermions. Thus, as an effective theory, we assume, in the forthcoming analysis, that the scattering amplitudes of the processes under investigation in this work can be computed in a perturbative series up to exotic quark masses of 1 TeV, and a more fundamental model will reveal a UV completion where new particles restore the unitarity of the color neutral $Q_1\bar{Q}_1\rightarrow Q_1\bar{Q}_1$ amplitude which, actually, is the more dangerous regarding the convergence of the perturbative series.

Anyway, even if this naive upper bound is valid, we are going to show that the 8 TeV LHC is able to search for $Y$ quarks well within the mass range where our perturbative computations apply. By the way, concerning the scattering amplitudes involved in $pp$ collisions at a hadron collider with initial state quarks and gluons, the perturbative limit is not violated even for very large exotic quark masses. Thus, in principle, the exotic quarks production cross sections can be computed reliably.

\section{Signals at a hadron collider}
\label{signals}

The production of the new particles predicted by this model lead to distinctive signals at a hadron collider. We list now some promising signatures.

\vskip0.5cm
\textbf{1-} \underline{$X$ quark pair production.}

\begin{eqnarray}
pp\rightarrow X\,\overline{X} & \rightarrow & t\, W^{-}\,\,\,\overline{t}\, W^{+}
\end{eqnarray}

This kind of process was studied, within the effective Lagrangian approach, in Refs.~\cite{contino-servante-2008,oscar}. Top quarks decays lead to a four $W$ final state and two hard bottom-jets. Requiring that at least two $W$ decays to a pair of jets and an opposite-sign lepton pair, it is possible to reconstruct the $X$-quark resonance but to a cost of a fourfold degeneracy due the escaping neutrinos and combinatorial backgrounds.

On the other hand, requiring that a $W^\pm W^\pm$ pair decays to leptons we have a very clean signal with two same-sign leptons. This channel has been investigated by the CMS collaboration~\cite{D0-exquark-2007a,D0-exquark-2007b,D0-exquark-2007c} in the search for exotic quarks that decay to top quarks and $W$ bosons. By the way, a lower bound of $\sim 650$ GeV has been placed for $5/3$ quarks from this analysis. We should point out that, in our model, the $X$ quark branching ratio into $tW$ depends upon the precise quark mixing so, in principle, it is possible to evade that limit. We postpone a detailed analysis on the other exotic quarks predicted by the model to a future investigation.

As we have mentioned in the Introduction, a quark with electric charge $5/3$  also arises within  the minimal 3-3-1 model~\cite{abd-2012}. In this case, the exotic quark is actually a fermionic leptoquark which a non zero lepton number, and whose decays lead to same-sign lepton pairs plus hard bottom jets.

\vskip0.5cm
\textbf{2-} \underline{$Y$ quark pair production.}

\begin{eqnarray}
pp\rightarrow Y\,\overline{Y} & \rightarrow & b\, W^{-}\,\,\,\overline{b}\, W^{+}
\end{eqnarray}

\begin{figure}
\centering
\includegraphics[scale=0.44]{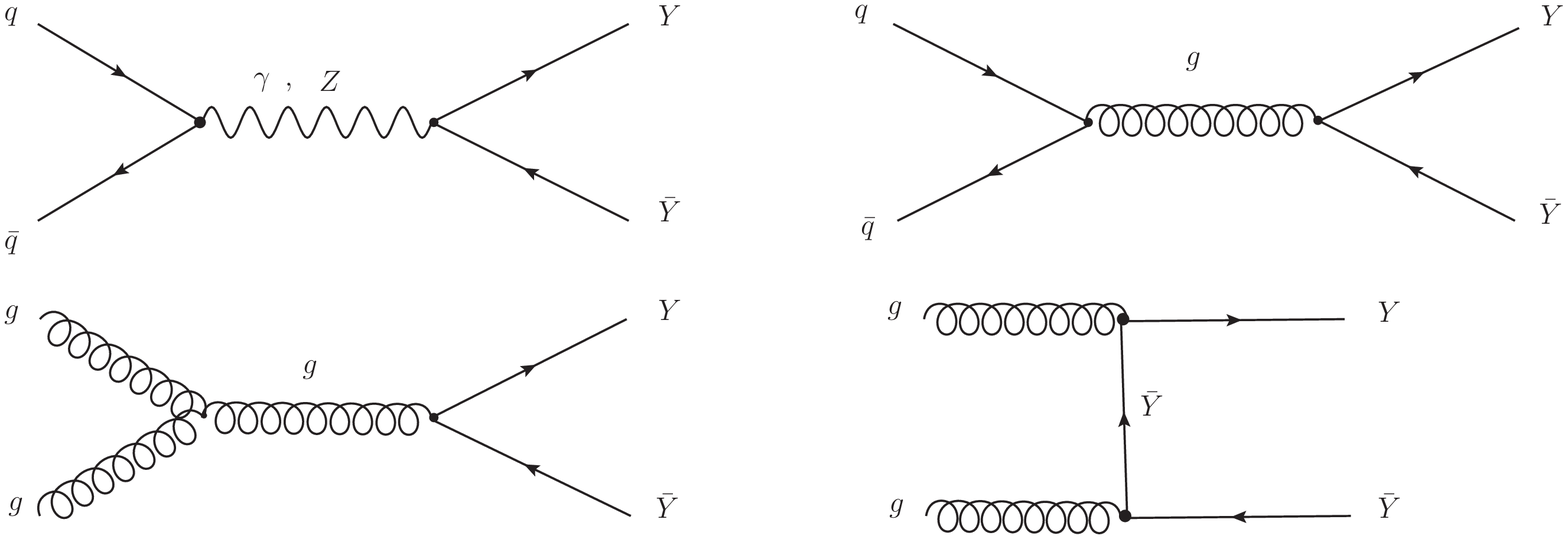}
\includegraphics[scale=0.44]{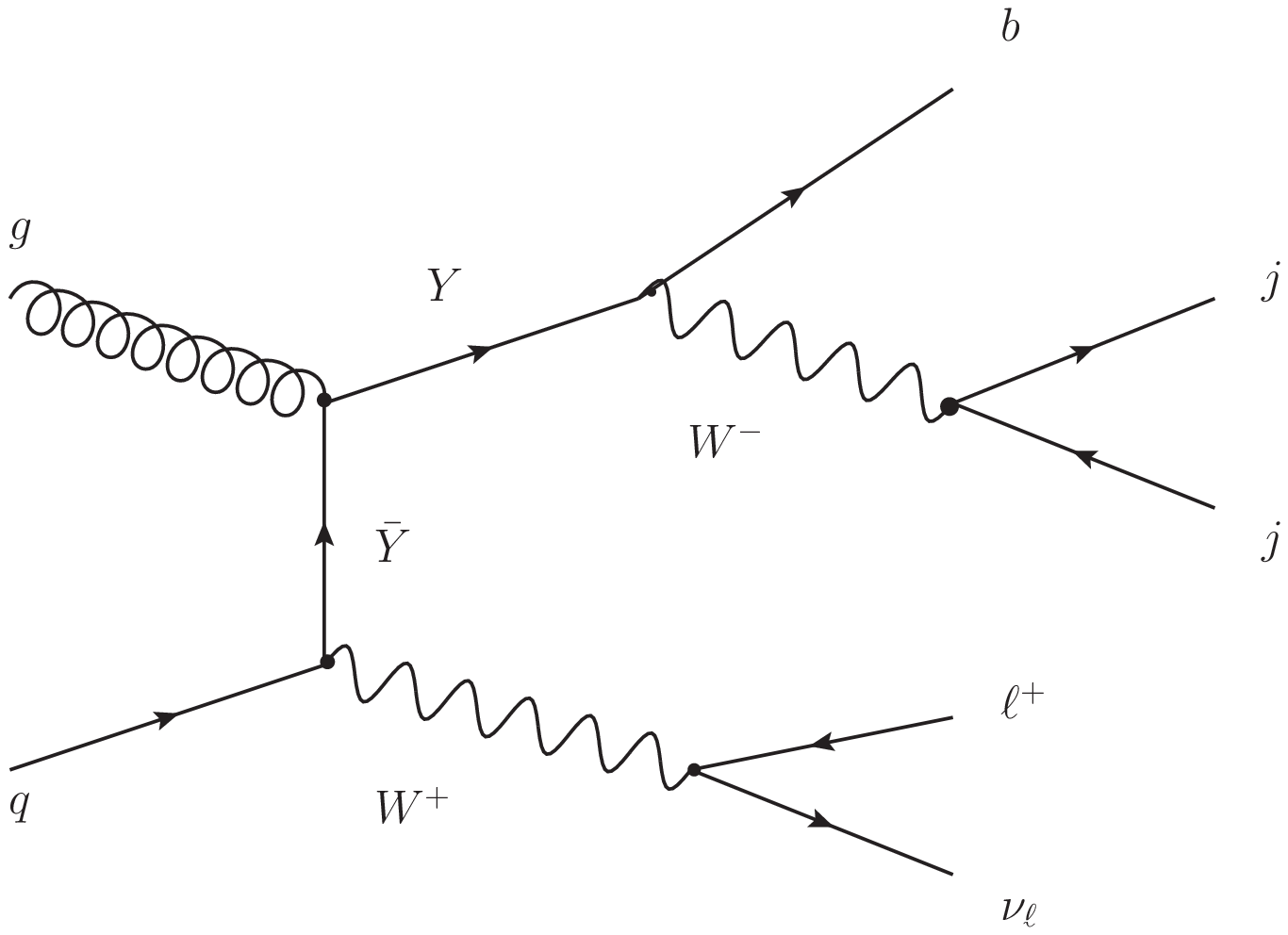}
\caption{Diagrams for pair and single production of the $Y$ quark at a hadron collider.}
\label{xyfeynman}
\end{figure}

In this case, the $Y$ quark is just a heavy top quark with a different electric charge.
As the $X$ quark, the production of a pair of $Y$ quarks proceeds through EW and QCD diagrams as shown in Figure \ref{xyfeynman}. It was pointed out in Ref.~\cite{cdf_charge} that the charges of $Y$ and $t$ quarks can be identified at a 99\% of confidence level in semileptonic final states by tagging the bottom-jet charge. Of course, this charge tagging technique can be used to eliminate the $Wb$ pairings from top quark decays.

\newpage
%\vskip0.5cm
\textbf{3-} \underline{$Y$ quark single production.}

\begin{eqnarray}
pp\rightarrow Y\, W^{+} & \rightarrow & b\, W^{+}W^{-}
\end{eqnarray}

This production process is proportional to the quark mixing matrix elements between the new $Y$ quark and the SM quarks (see the lower diagram of the Figure \ref{xyfeynman}).
The main background to this process is the single top quark production in association with a $W$ boson. Searches for an exotic quark of electric charge $4/3$, like the $Y$ quark, were done by the D0 Collaboration in \cite{D0-exquark-2007a,D0-exquark-2007b,D0-exquark-2007c,cdf_charge}. Signals of vectorial quarks of this charge were also investigated in Ref.~\cite{aguilar2009}. We perform a search analysis for a combined signal from single and pair $Y$ production at the 8 TeV LHC in the following section.

\vskip0.5cm
\textbf{4-} \underline{$Q_1Q_2$ production.}

 Another very interesting signal to search for is the pairs productions
\begin{equation}
du\rightarrow U D(XY)\;, \;\;\; uu \rightarrow XD\;, \;\;\; dd\rightarrow U Y
\end{equation}
through a t-channel $W,Z,h_1$. Beside these channels can lead to same-sign dileptons final states, their production cross sections are enhanced due the large luminosity of the initial state valence quarks which might compensate for the lack of strong interactions~\cite{buchkremer-et-al2013}.

\vskip0.5cm
\textbf{5-} \underline{$h_2$ Higgs boson production and decay.}

\begin{equation}
pp\rightarrow h_{2} \rightarrow \gamma\gamma,\; W^{+}W^{-},\; Z\, Z,\;\; F\overline{F},
\end{equation}
in which $F$ stands for either a new lepton $E_L^\pm$, $E_L^{\prime\pm}$, $F_L^{\pm\pm}$ or a new quark $X$, $Y$, $U$ and $D$, as long as the $h_2$ is heavier than the new fermions.

 As the Higgs boson $h_{2}$ comes out from a second Higgs doublet, which
couples only with the new fermions in the model according to our prior assumption,
its production will be dominated by the gluon fusion with the new quarks running in the loop. The $h_{2}$ coupling to the SM fermions is small and arises through its mixing
with $h_{1}$, a SM like Higgs boson.

In the scenario where all the new fermion fields are heavier than $h_2$, it decays mainly into pairs of gauge bosons, so it would lead to an excess in the current search channels of the SM Higgs boson at the LHC. On the other hand, once a new fermion channel is open it becomes the preferred decay channel. In this case, hard leptons, including signatures of same-sign dileptons if the $F_L^{\pm\pm}$ is open, plus missing energy could be a promising channel if the new leptons are lighter than $h_2$. Of course, new quarks could also be produced in the resonant $h_2$ production. A detailed analysis of the LHC potential for discovering $h_2$ will be done elsewhere.

\vskip0.5cm
\textbf{6-} \underline{Fermionic dark matter.}

If the new leptons are odd under some kind of discrete symmetry, then the neutral heavy lepton $N$, being the new lightest particle, would be a fermionic dark matter candidate. In this case, its connection to the SM sector would be through the $h_2$ couplings which constitute a resonant dark matter--Higgs portal scenario.

As the $h_2$ production through gluon fusion is expected to be large if the heavy Higgs is not too heavy, a hard monojet plus missing energy constitutes a smoking gun signature of dark matter production at the LHC in our model
\begin{equation}
pp\rightarrow h_2+g(q)\rightarrow N\overline{N}+j
\end{equation}

We will postpone a detailed study of the Higgs sector and the dark matter candidate of the model to the future but focus, in the next section, on the search for the $Y$ quark.

\section{Phenomenology of the $Y$ quark}
\label{pheno}
\subsection{Production cross section and total width}

In this section we present the prospects for the $Y$ quark search at the 8 TeV LHC. We focus on the $Y$ quark as it requires a more straightforward analysis compared to an $X$ quark which decays to top quark plus $W$ boson. The $X$ quark pair production would lead to four $W$ bosons and two bottom-jets which make the $X$ quark reconstruction more difficult. Moreover, a $Y$ quark was recently searched for in the Tevatron~\cite{exotic_charge1,exotic_charge2} in order to confirm or not the electric charge of the top quark. Due this experimental analysis we can safely start the search for $m_Y > m_t$.

\begin{figure}
\centering
\includegraphics[scale=0.5]{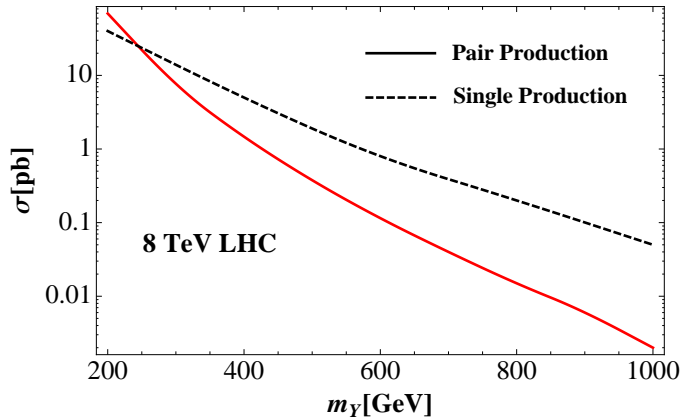}
\caption{Pair and single production cross sections for the $Y$ quark at the 8 TeV LHC.
%The mass span was chosen in order to anticipate the complementarity of the 8 TeV and 14 TeV runs in the search for the $Y$ quark.
}
\label{xsec}
\end{figure}

The $Y$ quark can be pair produced at the LHC in the $q\bar{q}$ and $gg$ annihilation channels as depicted in Figure~\ref{xyfeynman}. Although the EW contributions are enhanced due a larger electric charge, the QCD production contributes most to the pair production cross section. On the other hand, single production at order ${\cal O}(\alpha_s\alpha)$ may become important as the mass of the exotic quark gets heavier. This feature can be seen in Figure~\ref{xsec} where the single production dominates the cross section for masses above $\sim 300$ GeV at 8 TeV LHC.
%In the 14 TeV regime, the single production overwhelms the pair production for all masses from $800$ GeV onwards.
%

At this point we have made some assumptions about the mass matrix of the $d$-type quarks. First of all we assume that the $D$-quark mass is larger than the $Y$ quark mass in such way that the  $Y$ quark  decays into off-shell $D$ quark through the 3-body decay mode $Y\rightarrow D^*W^-\rightarrow q\,W^-W^-$ ($q=u,\,c,\,t$). Even though the $D$ and $Y$ quarks masses are just barely different, as required by the $STU$ parameters (see Figure \ref{fig-stu}), the $D$ quark decays to SM quarks are further suppressed by $|{\cal V}_{i4}|^2$ (see Appendix), so the $D^*(\rightarrow q+W^-)+W^-$ branching ratio is negligible.

Second, in order to evade flavor changing neutral currents constraints, the $({\mathcal{V}}^d_L)_{4d}$ and $({\mathcal{V}}^d_L)_{4s}$ mixing matrix elements have to be very small, then it is natural to expect a quark mixing pattern like $|({\mathcal{V}}^d_L)_{4d}|, |({\mathcal{V}}^d_L)_{4s}|\ll |({\mathcal{V}}^d_L)_{4b}|\ll |({\mathcal{V}}^d_L)_{4D}|$. As an outcome of these assumptions, the $Y$ quark is expected to decay almost exclusively to $b+W$. Moreover, the single production occurs through an initial state bottom quark from the proton sea: $bg\rightarrow Y+W$. This is surely a conservative scenario, once a $d$-quark initiated process would give a much higher single production cross section, despite some depletion from a lower branching ration into $b+W$ would occur. The cross sections shown in Figure~\ref{xsec} were calculated taking the benchmark scenario $({\mathcal{V}}^d_L)_{4b}=0.1$, $({\mathcal{V}}^d_L)_{4s}=0.008$, and $({\mathcal{V}}^d_L)_{4d}=0.005$.

The total width is given by the formula
\begin{equation}
\Gamma_{Y}= \frac{G_{F}}{8 \pi \sqrt{2}} m^{3}_{Y} \sum_{i=d,s,b} |({\mathcal{V}}^d_L)_{4i}|^{2},
\label{eq:width}
\end{equation}
in which it is considered that the $Y$ quark decays into the SM $d$, $s$ and $b$ quarks plus a $W$ boson, according to the assumptions we have just made. For the sum of the elements in the range $10^{-4}\leq\sum_i|({\mathcal{V}}^d_L)_{4i}|^{2}\leq 10^{-2}$, we have that for $m_Y=300$ GeV the width varies as $3.2 \times10^{-7}\,\hbox{GeV}\leq \Gamma_Y\leq 9\times 10^{-4}\,\hbox{GeV}$, and that for $m_Y=1$ TeV, $10^{-5}\,\hbox{GeV}\leq \Gamma_Y\leq 3.3\times 10^{-2}\,\hbox{GeV}$.
Because the $D^\prime$-quark channel is not open and given the smallness of the mixing parameters, the $Y$ quark widths are very narrow.

\subsection{Analysis and results}

It is possible to extend the discovery reach of the 8 TeV LHC up to a mass of $\sim 750$ GeV, assuming $BR(Y\rightarrow b+W)=100$\%, as we are going to show, by including events from single and pair production of $Y$ quarks. To do so we rely to a more inclusive analysis selecting the following semileptonic events
\begin{eqnarray}
& & pp\rightarrow Y\,W\rightarrow bW^-W^+\rightarrow bjj+\ell+\nu_\ell, \\
& & pp\rightarrow Y\overline{Y}\rightarrow b\bar{b}W^+W^-\rightarrow b\bar{b}jj+\ell+\nu_\ell,
\end{eqnarray}
where one of the $W$ bosons decays into a pair of jets and the other to a lepton $\ell$ (electron or muon) and a neutrino.

The signal and SM backgrounds events were simulated, assuming our benchmark scenario $(\mathcal{ V}^d_L)_{4b}=0.1$, $(\mathcal{V}^d_L)_{4s}=0.008$ and $(\mathcal{ V}^d_L)_{4d}=0.005$, with \texttt{MadGraph5}~\cite{mad5}, whereas the exotic quark model was implemented in the FeynRules~\cite{feynrules}. Hadronization and detector effects were taken into account from \texttt{Pythia}~\cite{pythia} and \texttt{PGS4}~\cite{pgs} interface to \texttt{MadGraph5} including the $k_T$-MLM jet merging scheme~\cite{mlm}.

We require events with, at least, three identified jets, one of them must be tagged as a b-jet, an isolated charged electron or muon, and missing energy. The b-tag efficiency is determined by the \texttt{PGS4} parametrization: $\sim 50$\% within $|\eta(b)|<1.0$ and decreased rates for $1.0<|\eta(b)|<2.0$ for a light quark/gluon mistag rate of a few percent for the same $\eta$ coverage. We checked that not requiring a tagged b-jet decreases significantly the discovery prospects of the 8 TeV LHC. This  part of the analysis that can be significantly improved using the most recent heavy flavor tagging techniques of the LHC collaborations.

The backgrounds considered in the analysis are: (1) the irreducible single top production, $tW^\pm$, (2) the irreducible top pair production, $t\bar{t}$, and (3) the main reducible ones: EW $t+j$, EW $t+b$, QCD and EW $jjj$ and $b\bar{b}j$.

As the top quark decays to $W^+b$ and the $Y$ quark to $W^-b$, it is possible to tag the top quark events by identifying the charge of the bottom quark in order to reconstruct the correct $Wb$ pairing~\cite{cdf_charge}. We did not try to apply this tagging technique, but found that a simple cut and count approach is able to separate the signals even though only a typical low acceptance rate has been achieved in this kind of analysis. For that  purpose we devise the following set of kinematic cuts
\begin{eqnarray}
p_{T_{j_{1(2)}}} &>& 40(20)\; \hbox{GeV}\;, \;\; p_T(b) > 200\; \hbox{GeV}\;, \;\; p_T(\ell) > 20\; \hbox{GeV}\label{cuts1a}  \\
|\eta(j)| &<& 2.5\;, \;\; |\eta(b)| < 2.5\;, \;\; |\eta(\ell)| < 2.5\label{cuts1b}  \\
\Delta R_{j_1j_2} &>& 0.4\;, \;\; \Delta R_{jb}>0.4\;, \;\; \Delta R_{j\ell}>0.4\;, \;\; \Delta R_{b\ell}>0.4 \label{cuts1c}\\
\not\!\! {E_T} &>& 100\; \hbox{GeV}\;, \;\; 60 < m_{j_1j_2} < 100\; \hbox{GeV}\;, \;\; m_{bjj} > m_Y-100\; \hbox{GeV}\label{cuts2}\\
H_T &>& 1\; \hbox{TeV}
\label{cuts3}
\end{eqnarray}

A very hard bottom-jet is expected from the decay of a heavy top partner so we include a hard $p_T(b)$ cut alongside the identification criteria of Eqs.~(\ref{cuts1a}), (\ref{cuts1b}), and (\ref{cuts1c}). After imposing the identification cuts, we require that a pair of jets, not tagged as b-jets, reconstruct the $W$ boson peak which effectively suppresses QCD and EW processes not related to a hadronic decaying $W$ in Eq.~(\ref{cuts2}). A hard missing transverse momentum cut eliminates the fake multi-jet QCD backgrounds.

\begin{figure}
\centering
\includegraphics[scale=0.6]{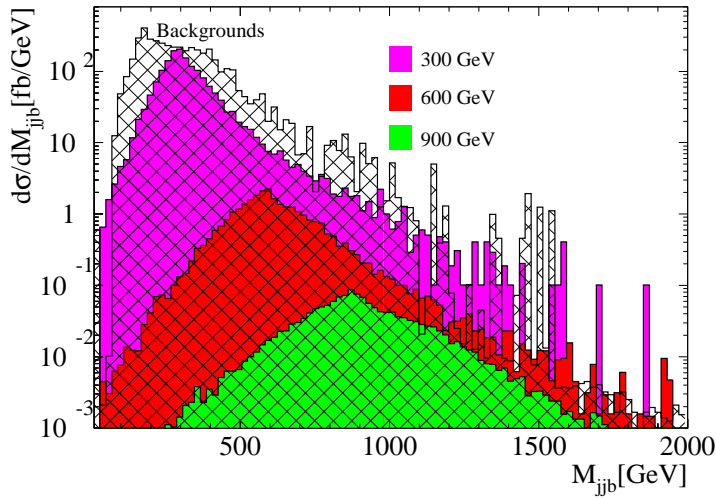}
\caption{\label{fig:dist} The $m_{jjb}$ invariant mass distribution for the signal ($300$, $600$, and $900$ GeV) and the total background. The variable presents harder events as the mass of the $Y$ quark gets larger.}
\end{figure}

The $m_{jjb}$ variable is constructed from the combination of the two jets, not tagged as a b-jet, and one tagged b-jet that minimizes the difference $|m_{jjb}-m_Y|$ for a given $Y$ quark mass. We checked that for a given mass assignment, the signal significance reaches a maximum when $m_{jjb}$ is close to the $Y$ mass and the cut value is somewhat smaller than $m_Y$.
The $jjb$ resonance can be explored in order to reject the $t\bar{t}$ and $tj$ backgrounds as can be seen in Figure~\ref{fig:dist}. As a rule of thumb, we found that cutting around $100$ GeV to the left of the resonance peak maximizes the signal significance for the masses considered.

The $H_T$ variable is another important discriminant between heavy exotic quarks and the SM backgrounds and comprises the scalar sum of all hadronic activity in the calorimeters plus the missing transverse energy. A hard cut in $H_T$, Eq.~(\ref{cuts3}), suppresses both the signal and backgrounds, but it is necessary to reach the discovery significance with a high $S/B$ ratio.

Even tough the actual values of the quark mixing matrix have a sizable impact on the production cross section, our quark mixing matrix is not phenomenologically optimistic at all. As long as the $Y$ quark decays exclusively to $b+W$ and its total width remains narrow, any changes in the $\left({\cal V}_L^d\right)_{4b}$ value would just rescale the single production rates and this change can be easily translated to a new signal significance. For that purpose we furnish the signal composition after imposing the kinematic cuts of Eqs.~(\ref{cuts1a}--\ref{cuts3}),
in the second column of the table~\ref{result}. The background cross section is constituted mainly by QCD $t\bar{t}$ and EW $t+j$ events after cuts. Assuming $23\; \hbox{fb}^{-1}$ of integrated luminosity at the 8 TeV LHC, it remains very few background events for this set of cuts, ranging from $0.6$ down to $0.3$ when $m_{jjb}>200$ GeV is hardened up to $m_{jjb}>800$ GeV.

As the number of expected background events is low, approximating a Poisson distribution with parameter $\lambda$ to a Normal distribution $N(\lambda,\sqrt{\lambda})$, in this case, overestimates the p-value associated to discovery for a given number of signal plus background events. The probability $\beta$ to accept the signal plus background hypothesis, when the background hypothesis is true, should be computed now with a Poisson PDF~\cite{cernyellowrep}
\begin{equation}
\beta = \sum_{n=\lfloor n_S+n_B\rfloor}^\infty \hbox{Poiss}(\lambda=n_B)
\end{equation}
where $\lfloor n_S+n_B\rfloor$ stands for the smaller integer greater than the theoretical number of signal plus background events from our simulations. We claim a discovery if $\beta < 2.8\times 10^{-7}$ which corresponds to a $5\sigma$ deviation from the mean of the standard Normal distribution.

We shown in the third column of the Table~\ref{result} the signal significance for a given $Y$ quark mass using the exact Poisson distribution to compute the p-values. We see that the 8 TeV LHC is able to discover this particle up to masses of $\sim 750$ GeV. A $3\sigma$ evidence can be reached for masses up to $\sim 850$ GeV.

 The fourth column shows the signal acceptance which reaches $1.4$\% for masses above $600$ GeV. Such low acceptance rates were found also in similar works on the search for exotic quarks~\cite{cacciapaglia-2011,cacciapaglia-2012}.

The maximum cross section occurs for a $300$ GeV $Y$ quark. As the mass of the $Y$ increases, its production cross section decreases as we see in the Figure~\ref{xsec}, however, as it gets heavier, its decay yields become harder and are likely to pass the cuts. For larger masses the acceptance is always increasing, but the production cross sections decrease steeply. Note also that the inclusion of the single quark contribution has a important impact for all masses, but the relative contribution becomes more relevant for heavier masses.

We also checked that our results and fairly insensitive to systematic uncertainties on the background total cross section up to $20$\% even for a $900$ GeV $Y$ quark. This is a consequence of the high $S/B$ ratios achieved after cuts, typically larger than $3$.
\begin{table}
\centering
\begin{tabular}{|c|c|c|c|c|}
\hline
  Mass (GeV)  & Signal (fb)  & Significance & Acceptance (\%) \tabularnewline
\hline
\hline
 300 & $5(1.3+3.7)$ & $31.2$ & $0.1$ \tabularnewline
\hline
 400 & $2.7(0.8+1.8)$ & $21.7$ & $0.2$ \tabularnewline
\hline
 500 & $2.2(0.7+1.5)$ & $19.4$ & $0.7$ \tabularnewline
\hline
 600 & $1.4(0.5+0.9)$ & $14.9$ & $1.4$ \tabularnewline
\hline
 700 & $0.5(0.2+0.3)$ & $7.8$ & $1.4$ \tabularnewline
\hline
 800 & $0.18(0.08+0.1)$ & $4.3$ & $1.3$ \tabularnewline
\hline
 900 & $0.07(0.03+0.04)$ & $2.1$ & $1.3$ \tabularnewline
\hline
\end{tabular}
\caption{The second column shows the signal cross sections after cuts and b-tagging for $Y$ quark masses from $300$ GeV up to $900$ GeV at the 8 TeV LHC. In parenthesis we quote the signal composition (single production+pair production). The third column presents the statistical significance achieved for $23\; \hbox{fb}^{-1}$. In the last column, the signal acceptance.}
\label{result}
\end{table}

Concerning the 14 TeV LHC, assuming an integrated luminosity of $100\; \hbox{fb}^{-1}$, around 10 signal events are expected for a $Y$ quark of mass $\sim 2$ TeV. Imposing a very hard cut on the $m_{jjb}$ invariant mass should effectively clean the background events, but only a detailed simulation including pile up effects also, would be able to evaluate the actual signal acceptances. Anyway, the 14 TeV LHC will extend the search region for $Y$ quarks complementing the searches of the 8 TeV run assuming that multi-TeV mass region is not departed from the perturbative regime of the theory.

%Concerning the 14 TeV LHC, we show in the Figure~\ref{xsec} the pair and single production cross section of $Y$ quarks. Assuming an integrated luminosity of $100\; \hbox{fb}^{-1}$, around 10 signal events are expected for a $Y$ quark of mass $\sim 2$ TeV. Imposing a very hard cut on the $m_{jjb}$ invariant mass should effectively clean the background events, but only a detailed simulation including pile up effects also, would be able to evaluate the actual signal acceptances. Anyway, the 14 TeV LHC will extend the search region for $Y$ quarks complementing the searches of the 8 TeV run.

\section{Conclusions}
\label{conclusao}

Given that the idea of quarks and leptons beyond the third generation is almost ruled out by the experimental evidences found at the LHC, including the recent discovery of a neutral scalar very similar to the SM Higgs boson, it
is natural to ask which might be the simplest way to extend the fermion sector of the SM. Some models propose new leptons and quarks which, in some cases, carry exotic electric charges. These exotic quarks and leptons have been searched for in the experiments at the Tevatron and the LHC.

In this paper, we present a simple theoretical construction that allows for new quarks and leptons, different from a SM sequential fourth family, with exotic electric charges. By considering the SM gauge group and the cancellation of anomalies we postulate a set of new chiral fermions with these features: quarks with electric charges $-4/3$ and $5/3$, singly and doubly charged leptons, and a heavy neutral lepton.

We have checked that the model respects all the constraints from the precision electroweak data, at the 95\% confidence level, evaluating the impact of the contribution of the new fermions to the $S$, $T$ and $U$ parameters. Flavor changing neutral currents arise in the model but they are suppressed by a high energy scale related to a more fundamental theory. This fundamental theory  is also assumed to restore the unitarity of heavy fermion scattering amplitudes in such a way that the $Y$ mass range, which is expected to be covered by the 8 TeV LHC run, is reliably predicted by our analysis. Anyway, even if the $600$ GeV upper bound is indeed valid, the 8 TeV LHC data is able to unravel a $Y$ quark with high significance. 

We also investigated the reach of the 8 TeV LHC to discover a $Y$ quark with electric charge $-4/3$. This quark decays similarly as a top quark but it can be much heavier. We found that considering both single and double $Y$ production, the 8 TeV LHC is able to discover a signal in
the $bjj+\ell+\not\!\!{E}_T$ channel, requiring at least one jet to be tagged as b-jet, for $Y$ masses from $300$ to $750$ GeV approximately. The 14 TeV run is expected to extend the search limits, although a high integrated luminosity will be probably required.

The model also predicts the existence of a heavy CP even neutral Higgs boson, $h_2$, which could have feeble interactions with SM fermions, and that might have a sizeable decay rate in pairs of gauge bosons. Thus, any hints of a second scalar in the Higgs boson search in $\gamma\gamma$, $ZZ$ and $WW$ channels would be a potential test to the model.

Imposing some kind of discrete symmetry on the lepton sector, at least, the new neutral lepton becomes stable and a fermionic dark matter candidate. This particles could communicate to the SM sector through the heavy Higgs boson $h_2$ and gives rise to a resonant dark matter--Higgs portal scenario.

\textbf{Note added:} As we finished this work the paper in Ref. \cite{abhv-2013} appeared with results on exotic vector-like quarks.

\acknowledgments
The authors acknowledges financial support from the Brazilian agencies FAPESP, under the process 2010/18762-6 (E.R.B), CNPq, under the process 302138/2010-2 (A.G.D), and  Universidade Federal do ABC (D.A.C). The authors also thank A. A. Natale for the suggestions in the manuscript, and Luca Panizzi and Mathieu Buchkremer for pointing out some interesting search channels.

\appendix
\section{Quarks mixing\label{sec:ApA}}

In order to establish the quarks mixing under the considerations in Section \ref{model},
we use the method of block diagonalization, frequently used in neutrinos physics,
which we learned from Refs.~\cite{lee-shrock1977,valle82,grimus-lavoura2000}. The mass matrix $\mathcal{M}^{q}$
in Eq. (\ref{lmass}) is turned into a diagonal form by a bi-unitary transformation
\begin{equation}
\mathcal{V}_{L}^{q\dagger}\mathcal{M}^{q}\mathcal{V}_{R}^{q}=
diag\left(m_{q1},m_{q2},m_{q3},m_{Q}\right).
\label{but}
\end{equation}
$\mathcal{V}_{L,R}^{q}$ links the symmetry eigenstates $\mathbf{q}_{L,R}^{\prime}$, to the mass eigenstates $\mathbf{q}_{L,R}\equiv\left(q_{1},q_{2},q_{3},Q\right)_{L,R}$, according to
\begin{align}
\mathbf{q}_{L}^{\prime} & = \mathcal{V}_{L}^{q}\mathbf{q}_{L}\nonumber \\
\mathbf{q}_{R}^{\prime} & = \mathcal{V}_{R}^{q}\mathbf{q}_{R}\,.
\label{bc}
\end{align}
We also define a generic $4\times4$ hermitian matrix
\begin{align}
\mathcal{M}^{q}\mathcal{M}^{q\dagger} & =\left[\begin{array}{cc}
\alpha_{3\times3}^{q} & \beta_{3\times1}^{q}\\
\beta_{1\times3}^{q\dagger} & \gamma_{Q}
\end{array}\right]\nonumber \\
 & =\left[\begin{array}{cc}
M^{q}\, M^{q\dagger} & M^{q}\, x^{*}\\
x^{T}M^{q\dagger} & x^{\dagger}x+\mid M_{Q}\mid^{2}
\end{array}\right]\label{mmd}
\end{align}
with the matrices $\alpha$, $\beta$, and the number $\gamma$ directly
defined in the last line of Eq. (\ref{mmd}). If the entries $\beta_{i}$
suppressed by a factor $v_1 v_2/\Lambda^{2}$, and the value of $\sqrt{\gamma}$
taken as a mass scale of the new quarks, there is a hierarchy for
the $\alpha$ and $\beta$ entries according to $\beta_{i}\ll\alpha_{ij}<\gamma$.

The matrix in Eq. (\ref{mmd}) is transformed in a diagonal form by
$\mathcal{V}_{L}^{q}$
\begin{equation}
\mathcal{V}_{L}^{q\dagger}\mathcal{M}^{q}\mathcal{M}^{q\dagger}\mathcal{V}_{L}^{q}
=diag\left(m_{q1}^{2},m_{q2}^{2},m_{q3}^{2},m_{Q}^{2}\right).
\label{mmdd}
\end{equation}
It is assumed that
\begin{equation}
\mathcal{V}_{L}^{q}=R^{q}\, U^{q}\label{WU}
\end{equation}
where the $4\times4$ matrix
\begin{equation}
R^{q}=\left[\begin{array}{cc}
\left(\sqrt{\mathbf{1}-B^{q}B^{q\dagger}}\right)_{3\times3} & B_{3\times1}^{q}\\
-B_{1\times3}^{q\dagger} & \sqrt{1-B^{q\dagger}B^{q}}
\end{array}\right]\label{Wm}
\end{equation}
transforms $\mathcal{M}^{q}\mathcal{M}^{q\dagger}$ into a block diagonal
form, and $U^{q}$ is a unitary matrix
\begin{equation}
U_{L}^{q}=\left[\begin{array}{cc}
\left(V_{L}^{q}\right)_{3\times3} & 0_{3\times1}\\
0_{1\times3} & 1
\end{array}\right]\label{Um}
\end{equation}
which turn $R^{q\dagger}\mathcal{M}^{q}\mathcal{M}^{q\dagger}R^{q}$
in a diagonal form. $V_{L}^{q}$ The square root in Eq. (\ref{Wm})
is defined as

\begin{equation}
\sqrt{\mathbf{1}-B^{q}B^{q\dagger}}\equiv\mathbf{1}-\frac{1}{2}B^{q}B^{q\dagger}
+\frac{1}{4}B^{q}B^{q\dagger}B^{q}B^{q\dagger}+...
\label{sqb}
\end{equation}
The matrix $B^{q}$ is a power series
\begin{equation}
B^{q}=B_{1}^{q}+B_{2}^{q}+...
\label{psB}
\end{equation}
in which  $B_{n}^{q}\sim \mathcal{O}\left(\left(\frac{v_{1}v_2}{\Lambda^2}\right)^{n}\right)$. In fact, it can be seen from the block diagonalization  that
\begin{equation}
B_{1}^{q}
\approx\gamma_{Q}^{-1}\beta^{q}\approx \frac{1}{\mid M_{Q}\mid^{2}}M^qx^*.
\label{b1}
\end{equation}
Thus, at first order in $B_{1}^{i}$ we have
\begin{equation}
\mathcal{V}_{L}^{q}=\left[\begin{array}{cc}
V_{L}^{q} & B_{1}^{q}\\
-B_{1}^{q\dagger}V_{L}^{q} & 1
\end{array}\right]\label{VL1}
\end{equation}
In this way the new quarks can decay into the ordinary quarks through
charged current interactions
\begin{equation}
\mathcal{L}^{c.c}=-\frac{g}{\sqrt{2}}\overline{\mathbf{u}}_{L}
\mathcal{V}\gamma^{\mu}\mathbf{d}_{L}W_{\mu}^{+}
+h.c.\label{cci}
\end{equation}
with the mixing matrix being
\begin{align}
\mathcal{V} & =\mathcal{V}_{L}^{u\dagger}\mathcal{V}_{L}^{d}\nonumber \\
 & \approx\left[\begin{array}{cc}
V_{CKM} & V_{L}^{u\dagger}\left(B_{1L}^{d}-B_{1L}^{u}\right)\\
\left(B_{1L}^{u\dagger}-B_{1L}^{d\dagger}\right)V_{L}^{d} & 1
\end{array}\right]\label{Vmix}
\end{align}
$V_{CKM}=V_{L}^{u\dagger}V_{L}^{d}$ is the usual Cabibbo-Kobayashi-Maskawa
matrix. At order $v_1v_2/\Lambda^2$, elements relevant for decay of exotic quarks $U^\prime$, $D^\prime$, $X$, and $Y$, in Eqs. (\ref{ccUd}), (\ref{ccuD}), (\ref{ccXu}), and (\ref{ccYd}), into ordinary quarks are then,
\begin{eqnarray}
& &\left(\mathcal{V}\right)_{4i}=\frac{1}{2\sqrt2}\frac{v_1v_2}{\Lambda^2}
\left[\left(\frac{v_2}{m_U^2}c^{uT} M^{u\dagger}-\frac{v_1}{m_D^2}c^{dT} M^{d\dagger}\right)V_{L}^{d}\right]_i,\cr
& &\left(\mathcal{V}\right)_{i4}=\frac{1}{2\sqrt2}\frac{v_1v_2}{\Lambda^2}\left[V_{L}^{u\dagger}
\left(\frac{v_1}{m_D^2} M^{d}c^{d*}-\frac{v_2}{m_U^2} M^{u}c^{u*}\right)\right]_i,\cr
& &\left(\mathcal{V}_{L}^{u}\right)_{4i}=-\frac{1}{2\sqrt2}\frac{v_1 v_2^2}{m_U^2\Lambda^2}\left(c^{uT} M^{u\dagger}V_{L}^{u}\right)_i,\cr
& &\left(\mathcal{V}_{L}^{d}\right)_{4i}=-\frac{1}{2\sqrt2}\frac{v_1^2v_2}{m_D^2\Lambda^2}\left(c^{dT} M^{d\dagger}V_{L}^{d}\right)_i,
\end{eqnarray}
with the order of the unity coefficients $c_i^q$ originating from Eq. (\ref{Qqmix}). In the presence of the singlet scalar field $\phi$, introduced in the end of Sec. \ref{model} we have effectively to change these formulas by replacing $\frac{1}{2}\frac{v_1v_2}{\Lambda^2}c_i^{u,d}\rightarrow \frac{1}{2}\frac{v_1v_2}{\Lambda^2}c_i^{u,d}+\frac{1}{\sqrt2}\frac{v_\phi}{\Lambda}a_i^{u,d}$, in which  we have the vacuum expectation value $\langle\phi\rangle=\frac{v_\phi}{\sqrt2}$.

%%%%%%%%%%%%

\end{document}